\documentclass[prl,aps,twocolumn,showpacs,superscriptaddress,%
floatfix]{revtex4}
\usepackage{graphicx}
\usepackage{bm}
\usepackage{amsmath}
\begin{document}
\title{Colossal magnetoresistance in manganites as 
a multicritical phenomenon}
\author{Shuichi Murakami}
\affiliation{Department of Applied Physics, University of Tokyo,
Bunkyo-ku, Tokyo 113-8656, Japan}
\author{Naoto Nagaosa}
\affiliation{Department of Applied Physics, University of Tokyo,
Bunkyo-ku, Tokyo 113-8656, Japan}
\affiliation{CERC, AIST Tsukuba Central 4, Tsukuba 305-8562, Japan}
\date{\today}
\begin{abstract}
The colossal magnetoresistance in manganites ${\it A}$MnO$_3$ 
is studied from the viewpoint of multicritical phenomena. To understand 
the complicated interplay of various phases,
we study the Ginzburg-Landau theory in terms of both the mean-field 
approximation and the renormalization-group analysis to 
compare with the observed phase diagram. 
Several novel features, such as the first-order ferromagnetic 
transition, and the dip in the transition temperature near 
the multicritical point, can be understood as driven by enhanced fluctuations 
near the multicritical point. Furthermore, we 
obtain a universal 
scaling relation for the $H/M$-$M^2$ plot (Arrott plot), which fits 
rather well with the experimental data, providing the further evidence for 
the enhanced fluctuation.
\end{abstract}
\maketitle
\begin{figure}
 \includegraphics[scale=0.5]{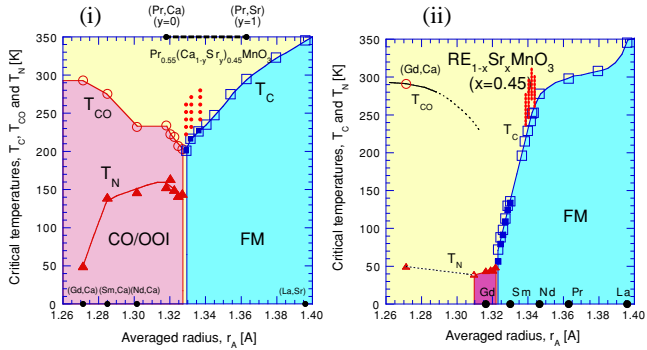}
 \caption{Phase diagram for the manganites: 
(i) Pr$_{0.55}$(Ca$_{1-y}$Sr$_{y}$)$_{0.45}$MnO$_{3}$ 
\cite{tt-i} and
(ii) (Nd$_{1-y}$Sm$_{y}$)$_{0.55}$Sr$_{0.45}$MnO$_{3}$ 
\cite{tt-ii}. 
Solid squares represent a first-order transition.
Red dots represent the data used in the scaling analysis in the text.
The red region at $0\text{K}<T<50\text{K}$ in (ii) is a spin-glass phase.}
 \label{phase}
 \end{figure}

Colossal magnetoresistance (CMR) in manganites is one of the most 
dramatic phenomena shown by strongly correlated electronic 
systems, and extensive experimental 
studies have revealed many aspects of this 
effect \cite{tokuranagaosa,dagotto}.
However, its mechanism has been the subject 
of long-standing debates; many theories have been 
proposed such as double exchange \cite{furukawa}, 
polaronic effect \cite{millis}, 
and phase separation combined with
percolation \cite{phaseseparation,phaseseparationt} and 
Griffiths singularity \cite{salamon}.
Shown in Figs.~\ref{phase} are the phase diagrams of the CMR manganites.
They clearly evidence that the CMR is related to the concomitant 
antiferromagnetic (AF) spin ordering,
charge ordering (CO), and orbital ordering (OO) around the
hole concentration $x=0.5$. 
Near the phase boundary between the
AF/CO/OO and the ferromagnetic metallic (FM) state, 
the transition temperature has a sharp dip and
the critical magnetic field $H_c$ is reduced considerably.
Hence, the CMR is 
collective in nature and differs from single particle
properties such as the transition from small to large polarons.

One of the subtle issues is the effect of the randomness on the CMR.
The ramdomness induced by the alloying and/or the nonuniform strain 
affects physical quantities near the boundary of the two phases, and alters
the critical properties easily. Figures \ref{phase} show the two systems (i)
Pr$_{0.55}$ (Ca$_{1-y}$Sr$_y$)$_{0.45}$MnO$_3$ and 
(ii) (Nd$_{1-y}$Sm$_y$)$_{0.55}$ Sr$_{0.45}$MnO$_3$.
It is not trivial which system is more disordered, 
namely alloying (i) alkaline-earth atoms or 
(ii) rare-earth atoms. There are, however, three reasons to believe
that 
the system (i) shows more intrinsic properties than (ii) which 
is mostly dominated by the disorder effect. 
The first 
is that the strong suppression of the 
ferromagnetic transition temperature and the appearance of the spin-glass type 
state in (ii) is well reproduced by recent model calculation in 
\cite{phaseseparationt}. 
The second is that the phase diagrams in ordered and disordered half-doped 
manganites 
\textit{Ln}$_{1/2}$Ba$_{1/2}$MnO$_3$ \cite{akahoshi,kageyama}
closely resemble Fig.~\ref{phase} (i) and (ii), respectively.  
The last one is that the 
scaling fit works almost perfectly in (i), while it does only 
in the limited region in (ii),
as we show in this Letter. 
In (ii), 
the region near the phase boundary is dominated by the 
disorder effect. 
An appealing 
scenario for this behavior is the phase separation 
and/or percolation of the conducting paths \cite{phaseseparationt}.
This assumes a mixture of metallic and insulating domains;
by applying magnetic field the metallic domains expand
to result in the CMR.
This is the static picture of the resistance network model
controlled by the magnetic field.
However, experiments on diffuse X-ray scattering and
Raman scattering  
revealed that intrinsic fluctuation is dynamic in 
nature \cite{tokuranagaosa}. 
Therefore, we need to take into account thermal 
and/or quantum fluctuations 
near the phase boundary.
It is well-known that fluctuation is enhanced near a 
multicritical point, where more than two orders collide
with each other.
In this Letter we propose a new scenario based on this idea
that the CMR originates from the enhanced fluctuation near the 
multicritical point, which is controlled by the small external
magnetic field.
We construct the phenomenological Ginzburg-Landau (GL)
model based on symmetry argument, and next give
a renormalization-group (RG) analysis for the multicritical
phenomena to compare with experiments.
This picture explains the scaling law 
and the first-order ferromagnetic phase transition 
as well as the enhanced sensitivity 
to the external magnetic field 
near the phase boundary.

The ordering pattern of AF/CO/OO is complicated with 
an enlarged unit cell. Several microscopic models for this 
state have been proposed \cite{co}. We employ here instead the 
phenomenological GL theory.
We classify possible terms in the free energy functional 
according to the symmetry of the order parameters. 
The relevant order parameters 
are those of ferromagnetism $\vec{M}$, 
antiferromagnetism $\vec{S}$,
charge ordering $\rho$, and orbital ordering $\vec{T}$.
Here, we discuss the dimensionality of each order parameter.
Both $\vec{M}$ and $\vec{S}$ are three-dimensional, while $\rho$ is scalar.
The orbital pseudo-vector $\vec{T}$ is originally three-dimensional
but in the presence of the Jahn-Teller interaction, which 
prefers real linear combinations of the two wavefunctions
$x^2-y^2$ and $3 z^2 - r^2$, it should be regarded as 
two-dimensional: ${\vec{T}} = (T_x,T_z)$.
The free-energy functional should be rotationally invariant
in the spin space, but not in the 
orbital pseudo-spin space. Hence, $\vec{M}$ and $\vec{S}$ should
appear in the form of ${\vec{M}}^2$ and ${\vec{S}}^2$.
In contrast, third-order terms in $\vec{T}$'s are allowed, because
$-\vec{T}$ is not iquivalent to $-{\vec{T}}$.
Linear terms in $\vec{T}$'s are generated by the 
coupling to the external uniaxial strain field.

Next we consider the wavenumbers of the order parameters.
From the spatial pattern of the AF/CO/OO,
we can easily see that wavenumber of each order is 
the following,
$\rho$: $(\pi,\pi,0 )$,
$\vec{S}$: $(\pi,0,\pi )$, $(0,\pi,\pi)$, $(\pi/2,-\pi/2,\pi)$, 
$(-\pi/2,\pi/2,\pi)$, 
$\vec{T}$: $(0,0,0)$, $(\pi,\pi,0)$, $(\pi/2,\pi/2,0)$, 
$(-\pi/2,-\pi/2,0)$, $\vec{M}$: $(0,0,0)$.
Therefore, the allowed terms in the 
GL functional are
\begin{eqnarray}
&&F=\frac{1}{2} \int d^3 r 
\biggl[ (\nabla {\vec{M}})^2 + (\nabla {\vec{S}} )^2 
 + (\nabla {\vec{T}} )^2 + (\nabla \rho )^2
\nonumber \\
&&\ \  +r_M ({\vec{M}} )^2 + r_S ({\vec{S}} )^2
+ r_T ({\vec{T}} )^2  + r_\rho (\rho )^2  
\nonumber \\
&&\ \  +( r_{\rho T}^x T_x  +   r_{\rho T}^z T_z  ) \rho
+g_{\rho S}\rho {\vec{S}}^2 + g_T \text{Re} ( T_x + i T_z )^3
\nonumber \\
&&\ \ +u_M (({\vec{M}} )^2)^2 + u_S (({\vec{S}} )^2)^2 + 
u_T (({\vec{T}} )^2)^2
 + u_{\rho} \rho ^4 \biggr]. 
\label{GL}
\end{eqnarray}
Here some remarks are in order.
First, the bilinear coupling between $\rho$ and $\vec{T}$
enforces that these two orders accompany each other.
This agrees with experiments.
Second, the third-order term in $\vec{T}$ make the 
transition first-order. However the magnitude of the
jump at the transition depends on the relative values of
$r_\rho$ and $r_T$; when $r_\rho\ll r_T$,
it becomes nearly second-order, while 
it is strongly first-order in the other limit.
Experimentally, the CO/OO transition in the narrow band-width side
is nearly second-order, which means that the transition is 
mainly driven by the CO. Hence, we neglect below the 
orbital ordering $\vec{T}$.
Third, the term $\rho {\vec{S}}^2$ 
prohibits the AF {\it without} the CO, namely
the CO occurs at a temperature higher than or as high as the AF.
This is a generic conclusion from the symmetry independent of 
microscopic mechanisms, and is consistent with 
the experiments.

 \begin{figure}
 \includegraphics[scale=0.5]{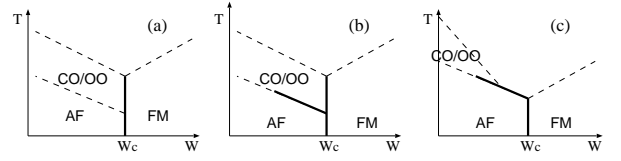}
 \caption{Mean-field phase diagrams for the GL functional (\ref{GL}) for 
in the $W$-$T$ plane. Here $W$ is the bandwidth and $T$ is the temperature.
The broken (solid) lines represent the second (first)-order phase
transitions.}
 \label{phase-GL}
 \end{figure}
Figure \ref{phase-GL} shows three possible mean-field phase diagrams 
for the GL functional (\ref{GL}), 
in the plane of the band-width $W$ and the temperature $T$.
Figure \ref{phase-GL}(a) is the most relevant to the 
experimental results. Nonetheless, this mean-field analysis cannot capture
several experimental features. One is the first-order FM
transition in the wide-bandwidth side.
Another is the dip of the transition temperatures of 
FM and CO/OO near the critical bandwidth $W_c$.
These two features are due to the fluctuations 
enhanced near $W_{c}$.
This is the so-called multicritical phenomenon between the
competing orders. 

We now turn to the RG
analysis of this fluctuation.
Let us consider a system with competing two orders ``A" and ``B".
We express their order parameters as an $N_{A}$-dimensional vector 
$\vec{\phi}_{A}$ and an $N_{B}$-dimensional vector $\vec{\phi}_{B}$.
Assuming a rotational invariance in $\vec{\phi}_{A}$  and
in $\vec{\phi}_{B}$, we can write
its classical free energy up to quartic order as
\begin{equation}
F_{{\text{cl}}}=
\frac{1}{2}(r_{A}\phi_{A}^{2}+r_{B}\phi_{B}^{2})+
\frac{1}{4}(u\phi_{A}^{4}+2w\phi_{A}^{2}\phi_{B}^{2}+
v\phi_{B}^{4}).
\label{Fcl}
\end{equation}
In the present case, the phases A and B correspond to the FM 
phase and the CO/OO phase. Thus, $\phi_{A}$ is a 
magnetization $\vec{M}$.
The average value $\frac{N_{A}r_{A}+N_{B}r_{B}}{N_{A}+N_{B}}$ is roughly 
proportional to the temperature measured from the multicritical point, 
while $r_{A}-r_{B}$ is proportional to the deviation of 
bandwidth from the multicritical point.
The model (\ref{Fcl}) was studied in 
\cite{nkf} by the RG. The fluctuation-induced 
first-order transition \cite{1storder} is known to occur in
(\ref{Fcl}) when $w>\sqrt{uv}$ \cite{mn}.
For an understanding of the fluctuation-induced first-order transition,
an effective potential \cite{amit} is a useful tool.
The effective potential is given by adding fluctuation contribution to the 
classical Lagrangian. Bare coefficients in the effective potential 
should be renormalized to remove 
ultraviolet divergence at $d=4$. By adding 
counterterms as in \cite{amit}, we get
\begin{eqnarray}
&&F=\frac{1}{2}(r_{A}\phi_{A}^{2}+r_{B}\phi_{B}^{2})+
\frac{1}{4}(u\phi_{A}^{4}+2w
\phi_{A}^{2}\phi_{B}^{2}+
v\phi_{B}^{4}) \nonumber \\
&&\ \ +(N_{A}-1)\frac{\alpha_{A}^{2}}{8}\left(\ln\frac{\alpha_{A}}{\kappa^{2}}
+\frac{1}{2}\right)
+\frac{\gamma_{+}^{2}}{8}\left(\ln\frac{\gamma_{+}}{\kappa^{2}}
+\frac{1}{2}\right)\nonumber\\
&&\ \ +(N_{B}-1)
\frac{\alpha_{B}^{2}}{8}\left(\ln\frac{\alpha_{B}}{\kappa^{2}}
+\frac{1}{2}\right)+\frac{\gamma_{-}^{2}}{8}
\left(\ln\frac{\gamma_{-}}{\kappa^{2}}
+\frac{1}{2}\right),
\label{eff}
\end{eqnarray}
where
\begin{eqnarray}
&&
\left(
\begin{array}{c}
\alpha_{A} \\
\alpha_{B}
\end{array}\right)
=\left(
\begin{array}{c}
  r_{A} \\  r_{B}
\end{array}\right)
+\left(
\begin{array}{cc}
u & w \\
w & v
\end{array}\right)
\left(\begin{array}{c}
\phi_{A}^{2} \\
\phi_{B}^{2}
\end{array}\right)
,\\
&&
\left(\begin{array}{c}
\beta_{A} \\
\beta_{B}
\end{array}\right)
=
\left(\begin{array}{c}
  r_{A} \\
 r_{B}
\end{array}\right)
+
\left(\begin{array}{cc}
  3u & w \\
  w & 3v
\end{array}\right)
\left(\begin{array}{c}
\phi_{A}^{2} \\
\phi_{B}^{2}
\end{array}\right),\\
&&
  \gamma_{+}+\gamma_{-}=\beta_{A}+\beta_{B}, \ \ 
  \gamma_{+}\gamma_{-}=\beta_{A}\beta_{B}-4w\phi_{A}^{2}\phi_{B}^{2},
\end{eqnarray}
and $\kappa$ is a parameter setting a scale of momentum.
We neglected the terms higher than $O(4-d)$.
The quantities $r_{i}$, $u$, $v$, $w$ in (\ref{eff}) are renormalized 
ones and are finite.

Let us consider the fluctuation-induced first-order 
transition \cite{1storder} between the disordered phase and the ``A" phase.
We thus assume that the ``B" field $\phi_{B}$ is not ordered, i.e.
$\phi_{B}=0$.
In the RG language, the fluctuation-induced first-order 
transition occurs when the RG flow runs into an unstable 
region of the model (\ref{Fcl}).
It means that sixth-order terms, though omitted in (\ref{Fcl}), 
are necessary for stability, leading to a first-order 
transition.
Hence, if the RG flow crosses the boundary 
of the stability region, $u=0$, the system undergoes a first-order 
transition to the ``A" phase.
In other words, we follow the RG flow for 
$u$, $v$, $w$ \cite{nkf}
until we reach the line $u=0$. Let $\kappa_{1}$ denote the value of 
$\kappa$ when
$u(\kappa)=0$.
Other quantities as $r_{A},r_{B},\phi_{A}$ are also renormalized.
It is worth noting that these quantities 
are renormalized multiplicatively, i.e.
$r_{i}(\kappa_{1})=r_{i}f(\kappa_{1}),\ 
\phi_{A}(\kappa_{1})=\phi_{A}g(\kappa_{1})$,
where $r_{i}$ and $\phi_{A}$ are the initial values.
Thus, renormalization is merely a change of scale for them.
The condition $u(\kappa_{1})=0$ greatly simplifies the above free energy as
\begin{equation}
  F=\frac{1}{2}r_{A}\phi_{A}^{2}+N_{A}f(r_{A})+N_{B}f(r_{B}+w
\phi_{A}^{2}),
\end{equation}
where $f(x)=~\frac{x^2}{8}\left(\ln x +\frac{1}{2}
\right)$ 
and $\kappa_{1}$ is set as unity since it can be absorbed by
the change of scale of other variables.
The equation of state is
\begin{equation}
H_{A}=\frac{\partial F}{\partial\phi_{A}}=r_{A}\phi_{A}
+2\phi_{A}N_{B}wf'(r_{B}+w\phi_{A}^{2}).
\label{hm}
\end{equation}
where $H_A$ is a field conjugate to the $\phi_{A}$
field.
Since we identify this $\phi_{A}$ as a magnetization $\vec{M}$, $H_{A}$ is
a magnetic field $H$.
It is convenient to rewrite (\ref{hm}) as
\begin{equation}
\frac{H}{M}=r_{A}
+2wN_{B}f'(r_{B}+wM^{2}).
\label{arrott}
\end{equation}
Therefore if we plot $H/M$ versus $M^2$ at various temperature and
bandwidth, all the plots will be degenerate.
In other words, the $H/M$-$M^2$ curve, called the Arrott plot, 
will undergo parallel transport
when we change temperature or bandwidth.
One would notice that in a certain range of temperature and bandwidth this 
universal curve crosses the horizontal axis once or twice. In that case some 
part of the universal curve becomes unphysical, and the system undergoes a 
first-order phase transition by increasing the magnetic field. The system 
is either ferromagnetic or metamagnetic.

To verify this scenario, we used two series of data in Figs.~\ref{phase}:
(i) for $y=0.25,0.3,0.4$ 
and (ii) for $0.1\leq y\leq 0.4 $,
with varying $y$ and $T$. We can regard $y$ as a parameter controlling 
the bandwidth. We expect that 
$r_{A}$ and $r_{B}$ are functions of $y$ and $T$ while $w$ is 
a constant. We expand $r_{A}$ and $r_{B}$ in the vicinity of the 
multicritical point as
$r_{A}=c_{AT}\Delta T+ c_{Ay}\Delta y$, 
$r_{B}=c_{BT}\Delta T+ c_{By}\Delta y$,
where $c_{AT},c_{BT},c_{Ay},c_{By}$ are constants.
In view of (\ref{arrott}), 
$c_{AT},c_{BT}$ ($c_{Ay},c_{By}$) represents an amount of parallel 
displacement of the $H/M$-$M^{2}$ plot when the temperature 
$T$ (doping $y$) is changed. 
We fitted the data in the following way. First we varied $c_{AT}, 
c_{BT},c_{Ay},c_{By}$ so that the plots for various $y$ and 
$T$ overlap most after the parallel displacement.
Then the scales of the abscissa and the ordinate are varied to fit to
(\ref{arrott}). 
We used the data shown as red dots 
in Figs.~\ref{phase}.
We discarded the data with $M>1.6\mu_{\text{B}}$, because when 
the magnetization 
approaches the saturation value, the GL functional up to quadratic 
order is no longer appropriate. 
We also discarded the data for $0.5\leq y\leq 0.8$ in 
(ii). They
do not fit well with the scaling curve.
This is reasonable because in (ii)
critical fluctuations are washed out near the multicritical point, 
by which the scaling plot deviates from (\ref{arrott}). 

\begin{figure}
 \includegraphics[scale=0.65]{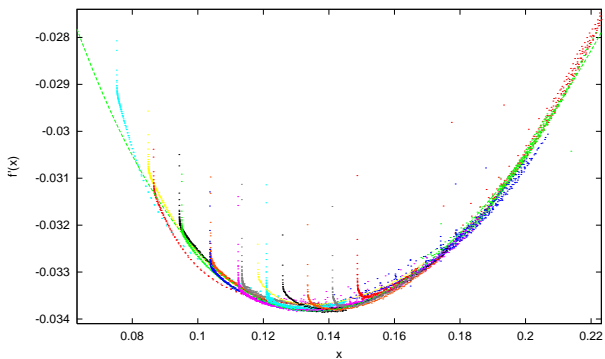}
 \caption{Scaling plot for 
(i) Pr$_{0.55}$(Ca$_{1-y}$Sr$_{y}$)$_{0.45}$MnO$_{3}$ and
(ii) (Nd$_{1-y}$Sm$_{y}$)$_{0.55}$Sr$_{0.45}$MnO$_{3}$. The 
broken line represent the $f'(x)=\frac{x}{4}(\ln x +1)$ 
plot. Each dot show the data
for fixed $T$ and $y$.}
 \label{plot}
 \end{figure}

The resulting plot is shown in
Fig.~\ref{plot}.  
The plots fall into one curve with good accuracy.
This shows that the fluctuation is enhanced near the multicritical point.
Because the critical scaling (\ref{arrott}) holds for a wide range of 
data shown in Fig.~\ref{plot}, the critical region is rather large 
($\sim 80\text{K}$). It manifests strong correlation of electrons in 
the manganites. 
Whether the system is critical or not is determined by 
$\xi$, measured by a length scale 
$\xi_{0}\sim \frac{E_{\text{F}}}{\Delta}a$, where $\Delta$ is a gap, $a$
is a lattice constant, and $E_{\text{F}}$ is the Fermi energy. 
In the manganites $\Delta\sim E_{\text{F}}$ yields 
$\xi_{0}\sim a$. 
Temperature dependence of $\xi$ in a related compound
Pr$_{0.5}$Ca$_{0.5}$MnO$_{3}$ can be obtained from Fig.~2(c) of 
\cite{shimomura}. It is $\xi\sim 100$\AA, near the transition temperature 
$T_{c}=235$K,
and is $\xi\sim 20${\AA} even at 300K, which is 
65K higher than $T_{c}$. Therefore,
$\xi$ is much longer that $\xi_{0}$ in the wide temperature range,
implying that the critical region should be large. 

The enhanced fluctuation makes the system sensitive to the
external magnetic field $H$. The CO/OO state easily becomes
the FM state by a weak magnetic field. Let $H_{c}$ denote this critical field.
An exponent $x$ defined by $H_{c}\propto (W_{c}-W)^{x}$ represents 
the sensitivity to the external field. 
Without the fluctuation, $x$ should be equal to unity.
On the other hand, in the multicritical region
but not in the fluctuation-induced first-order-transition region, $x$
is controlled by the bicritical fixed 
point, and is larger than unity. 
Calculation based on the $(4-\epsilon)$-expansion up to 
$O(\epsilon^{2})$ results in $x=\frac{\beta\delta}{\phi}\sim 1.37$,
where $\phi$ is the crossover exponent. 
It can be compared with the experimental value $x\sim 1.6$, extracted 
from the data for
Pr$_{0.65}$(Ca$_{1-y}$Sr$_{y}$)$_{0.35}$MnO$_{3}$ \cite{tomioka}. 
This increase of $x$
near the multicritical point, i.e. sensitivity to the external field, emerges 
as the CMR phenomena.

Here we discuss the effect of disorder, and 
the relevance of the Griffiths singularity \cite{salamon} to the present study.
It has been pointed out that in the temperature region $T_c< T<T_c^{(0)}$
where $T_c^{(0)}$ is the transition temperature for the pure system while 
$T_c$ is the suppressed one by the random dilution, there appears some 
singularity in the time-dependence of correlation 
functions of the order parameter,
such as the extended power-law decay. However, this singularity does not occur
in the thermodynamic quantities which we discussed in this Letter.
It is possible that the fixed point for the pure system is unstable against 
disorder, but according to the Harris criterion the bicritical fixed point
described above is at least locally stable \cite{amit}.
Hence, the thermodynamic properties are described by the usual bicritical 
fixed point, although the location of the critical point is shifted.
However, the disorder effect on the first-order phase transition 
are beyond the scope of the present work, and
ref.~\cite{phaseseparationt} is relevant to this issue.
The percolation scenario could also lead to the CMR, where 
the resistivity will depend sensitively on  
random realization of the metallic paths etc., and consequently on  
samples. This behavior has been actually observed in dilutely Cr-doped 
manganites \cite{cr}. Cr ions destroy the CO/OO locally, and introduce the 
FM region. In these dilutely doped samples, the resistivity at low temperature 
depends on the Cr-concentration dramatically, and the hysteresis 
appears in the temperature cycle. In contrast, when the Cr-concentration 
increases, the resisivity no longer depends on samples or heat cycle. 
In the latter case,  the thermodynamic phases are well-defined and the CMR
is triggered by the phase change between them, which is the 
subject of our study here. Hence, there are two types of the CMR; 
one is due to the percolating path and the other is due to the 
multicritical fluctuation near the phase change.
In the zero temperature limit, dynamics and 
statics are coupled and the Griffiths singularities influence the whole
quantum critical phenomena. In the manganites, this 
possibility seems to be prevented by the glassy state appearing at low 
temperatures.

We would like to acknowledge Y.~Tokura, Y.~Tomioka, E.~Dagotto, 
and H.~Kageyama 
for fruitful discussions.
This work is supported by Grant-in-Aids from the Ministry of Education, 
Culture, Sports, Science, and Technology.

\end{document}